\documentclass[aps,reprint,amsmath,amssymb, reprint,superscriptaddress,nofootinbib,floatfix]{revtex4-2}

\usepackage{graphicx}
\usepackage{dcolumn}
\usepackage{bm}
\usepackage{braket}
\usepackage{graphicx}
\usepackage{subfigure}
\usepackage{overpic}
\usepackage{dcolumn}
\usepackage{bm}
\usepackage{soul}
\usepackage{xr-hyper}
\usepackage[hidelinks,colorlinks=true,linkcolor=blue,citecolor=blue]{hyperref}
\usepackage[mathlines]{lineno}
\usepackage[title,titletoc]{appendix}
\usepackage{titlesec}
\usepackage{braket}
\usepackage{color}
\usepackage{xcolor}
\usepackage{float}
\usepackage{soul}
\usepackage{makecell}
\usepackage{amsthm}
\usepackage{subfiles}

\usepackage{titlesec}

\titleformat{\section}[runin]
  {\normalfont\normalsize\itshape}
  {Appendix \thesection}
  {0.6em}
  {}
  [.\ ---\hspace{0.5em}] 
\definecolor{grey}{rgb}{0.5, 0.5, 0.5}

\theoremstyle{definition}

\theoremstyle{remark}
 
\begin{document}
\preprint{}

\title{High-Efficiency Loading of 2400 Ytterbium Atoms in Optical Tweezer Arrays}

\author{Jiawen Zhu}
\altaffiliation{These authors contributed equally to this work.}
\affiliation{State Key Laboratory for Artificial Microstructure and Mesoscopic Physics and Frontiers Science Center for Nano-optoelectronics, School of Physics, Peking University, Beijing 100871, China}

\author{Changfeng Chen}
\altaffiliation{These authors contributed equally to this work.}
\affiliation{State Key Laboratory for Artificial Microstructure and Mesoscopic Physics and Frontiers Science Center for Nano-optoelectronics, School of Physics, Peking University, Beijing 100871, China}

\author{Li Zhou}
\affiliation{State Key Laboratory for Artificial Microstructure and Mesoscopic Physics and Frontiers Science Center for Nano-optoelectronics, School of Physics, Peking University, Beijing 100871, China}

\author{Xiangru Xie}
\affiliation{State Key Laboratory for Artificial Microstructure and Mesoscopic Physics and Frontiers Science Center for Nano-optoelectronics, School of Physics, Peking University, Beijing 100871, China}

\author{Chenyang Jiang}
\affiliation{State Key Laboratory for Artificial Microstructure and Mesoscopic Physics and Frontiers Science Center for Nano-optoelectronics, School of Physics, Peking University, Beijing 100871, China}

\author{Zhuoli Ding}
\affiliation{State Key Laboratory for Artificial Microstructure and Mesoscopic Physics and Frontiers Science Center for Nano-optoelectronics, School of Physics, Peking University, Beijing 100871, China}

\author{Fan Wu}
\affiliation{State Key Laboratory for Artificial Microstructure and Mesoscopic Physics and Frontiers Science Center for Nano-optoelectronics, School of Physics, Peking University, Beijing 100871, China}

\author{Fan Yang}
\affiliation{Hefei National Laboratory, Hefei 230088, China}

\author{Guoqing Wang}
\affiliation{International Center for Quantum Materials, School of Physics, Peking University, Beijing 100871, China}

\author{Qihuang Gong}
\affiliation{State Key Laboratory for Artificial Microstructure and Mesoscopic Physics and Frontiers Science Center for Nano-optoelectronics, School of Physics, Peking University, Beijing 100871, China}
\affiliation{Hefei National Laboratory, Hefei 230088, China}
\affiliation{Collaborative Innovation Center of Extreme Optics, Shanxi University, Taiyuan 030006, China}
\affiliation{Peking University Yangtze Delta Institute of Optoelectronics, Nantong, Jiangsu 226010, China}

\author{Peng Zhang}
\affiliation{School of Physics, Renmin University of China, Beijing 100872, China}
\affiliation{Key Laboratory of Quantum State Construction and Manipulation (Ministry of Education), Renmin University of China, Beijing 100872, China}

\author{Sheng Zhang}
\email{sheng.physik@pku.edu.cn}
\affiliation{State Key Laboratory for Artificial Microstructure and Mesoscopic Physics and Frontiers Science Center for Nano-optoelectronics, School of Physics, Peking University, Beijing 100871, China}

\author{Pai Peng}
\email{pengpai@pku.edu.cn}
\affiliation{State Key Laboratory for Artificial Microstructure and Mesoscopic Physics and Frontiers Science Center for Nano-optoelectronics, School of Physics, Peking University, Beijing 100871, China}
\affiliation{Collaborative Innovation Center of Extreme Optics, Shanxi University, Taiyuan 030006, China}

\begin{abstract}

Neutral atom arrays have emerged as a powerful platform for quantum computation, simulation, and metrology.
Among them, alkaline-earth-like atoms exhibit distinct advantages, including long coherence time, high-fidelity Rydberg gates, and erasure correction for efficient quantum error correction. However, their scalability has lagged behind that of the alkali atoms. 
Here, we report 2400 ytterbium-174 atoms trapped in an optical tweezer array with enhanced loading efficiency of 83.5(1)\% via blue-detuned light-assisted collisions. 
We develop a quantitative model of the collision dynamics and find good agreement between the calculated inelastic collision rates and the experimentally measured loading efficiencies.
Notably, the loading efficiency is largely maintained for array sizes ranging from dozens to thousands, exhibiting excellent scalability.
We further demonstrate that the enhancement exists robustly across a range of interatomic potentials, suggesting its utility for other atomic species.
To establish the capability of the $^{174}$Yb arrays toward universal quantum computation, we propose to encode the qubit in the ground-clock state manifold and estimate a 99.9\% two-qubit gate fidelity with experimentally feasible parameters.
Our work advances the prospects for realizing large-scale quantum computers using alkaline-earth-like atoms.
\end{abstract}

\maketitle
\textit{Introduction.---}
Neutral atoms in optical tweezer arrays have developed rapidly over the past decade, emerging as one of the most promising platforms for quantum information science \cite{lukin2016science-rearrange, AntoineBrowaeys2016science-rearrange, Korea2016NC-rearrange}. 
The rapid advancement is largely driven by their exceptional scalability, featuring thousands of atomic qubits with high-fidelity controls~\cite{2024PRX-1225atom, MPQ2024PRR-cloading, lukin2025nature-3000atom, 2024PRX-1225atom, endres2025nature-6100atom, ustc2025PRL-2024atom, AntoineBrowaeys2024prap-cryorearrange, Will2024arXiv-1300Sr,Birkl2024Optica-Rb}. 
Recently, significant progress has been made toward various applications.
For quantum computation, major milestones include the realization of high-fidelity Rydberg gates~\cite{jeff2025PRX-gatefidelity,lukin2023nature-gatefidelity,endres2025PRX-gatefidelity, kaufman2025arXiv-highfidelityentanglementcoherentmultiqubit, Atomcomputing2025prx-171gsnqubit,saffman2025PRX-individualaddress} and the demonstration of key fault-tolerant quantum computing elements \cite{lukin2024nature-logicalqubit,lukin2025nature-universal, Jeff2025arXiv-leaverageErasureError,Atomcomputing2024arxiv-fault}. 
As programmable quantum simulators, atom arrays offer unprecedented access to strongly correlated many-body phenomena, from quantum criticality to exotic states like quantum spin liquids~\cite{Normanyao2025Science-simulationtwz,AntoineBrowaeys2025science-simulation, ChenZhai2025PRL-Kibble-Zurek, LiLin2024arXiv-QIcollapse-revival,Lukin2021Science-spinliquid,youli2025PRL,chencheng2025Nature}. 
For precision metrology, the platform opens new frontiers for optical atomic clocks via quantum enhancement~\cite{kaufman2024nature-clock,endres2025nature-gatefidelity}. Furthermore, continuous innovations in the system architectures are unlocking new capabilities, such as coherent atom transport for nonlocal gates ~\cite{lukin2022Nature-nonlocalgate}, dual-species arrays for low-crosstalk control~\cite{zhanmingshen-mixed, hannes2022PRX-dualspecies, hannes2024NP-dualspecies}, local optical gates for faster control~\cite{ saffman2022nature-individualaddress, jeff2024optica-localgate, pengxu2025nc-fiberarray, zouchanglin2026scp-volcano, wangguoqing2025PRA-individualaddress,Drik2025NC-PIC}, efficient interface to photons~\cite{jacob2025np-network,Jacobcovey2023npj-network, Gephard2024Science-tweezernetwork,Lukin2025science-errorcavity, Stamper-Kurn2022PRL-midcavity, Simon2025arXiv-cavityarraymicroscopeparallel,Zhangtiancai2023PRL-cavity,wangguoqing2025arXiv-cavity}, and continuous reloading for rapid atom replenishment~\cite{lukin2025nature-3000atom, li2025fast,Bloch24prl-10000, MPQ2024PRR-cloading,2024PRX-1225atom}. 
These advances establish neutral atom arrays as a versatile and powerful platform for exploring the frontiers of quantum science and technology.

Among various atomic species, alkaline-earth-like atoms with two valence electrons, such as ytterbium and strontium, offer unique advantages for quantum control
. Pure nuclear spins provide excellent coherence for quantum information storage due to their isolation from environmental noise~\cite{kaufman2022PRX-enhanceloading,jeff2022PRX-UniversalGate}. The metastable states form the cornerstone of optical clocks, one of the most precise tools in quantum metrology~\cite{kaufman2020nature-twzclock,kaufman2024nature-clock,endres2025nature-gatefidelity}. Qubits encoded in the metastable state enable robust single-photon Rydberg excitation~\cite{endres2020np-AEAs,endres-nature2024benchmarking,jeff2023nature-erasure}, optical clock-state control~\cite{kaufman2023nature-opticalcontrol,kaufman2022NP-opticalcontrol,kaufman2023PRX-omg} 
, and midcircuit erasure conversion, which significantly reduces the resource overhead for quantum error correction~\cite{jeff2022nc-erasure,jeff2023nature-erasure, Jeff2025arXiv-leaverageErasureError,Jeff2023PRX-erasure,Endres2023Nature-erasure}.
However, an important challenge for alkaline-earth-like atom arrays is their limited size, as compared to alkali species with a record of 6100 atoms~\cite{endres2025nature-6100atom}. 
This is due to the low polarizability of alkaline-earth-like atoms at near-infrared wavelengths where high-power lasers are commercially available, therefore limiting the maximum number of optical tweezers with sufficient depth.
Moreover, for an optical tweezer array of fixed size, the $\sim 50\%$ statistical single-atom loading efficiency imposes a stringent constraint on the number of trapped atoms.
Blue-detuned light-assisted collision has been employed to enhance the loading efficiency for certain atomic species and array sizes up to 100~\cite{2015-enhanceloading,2010np-enhanceloading,2019PRX-enhanceloading,kaufman2022PRX-enhanceloading,Aliyu2021prr-enhanceloadingna23,Bryce2022prr-enhanceloadingk39}. Up to now,
it has been an open question whether the method can be broadly applied to different species or much larger arrays.

\begin{figure*}[!htbp] 
\centering 
\includegraphics[width=0.9\linewidth]{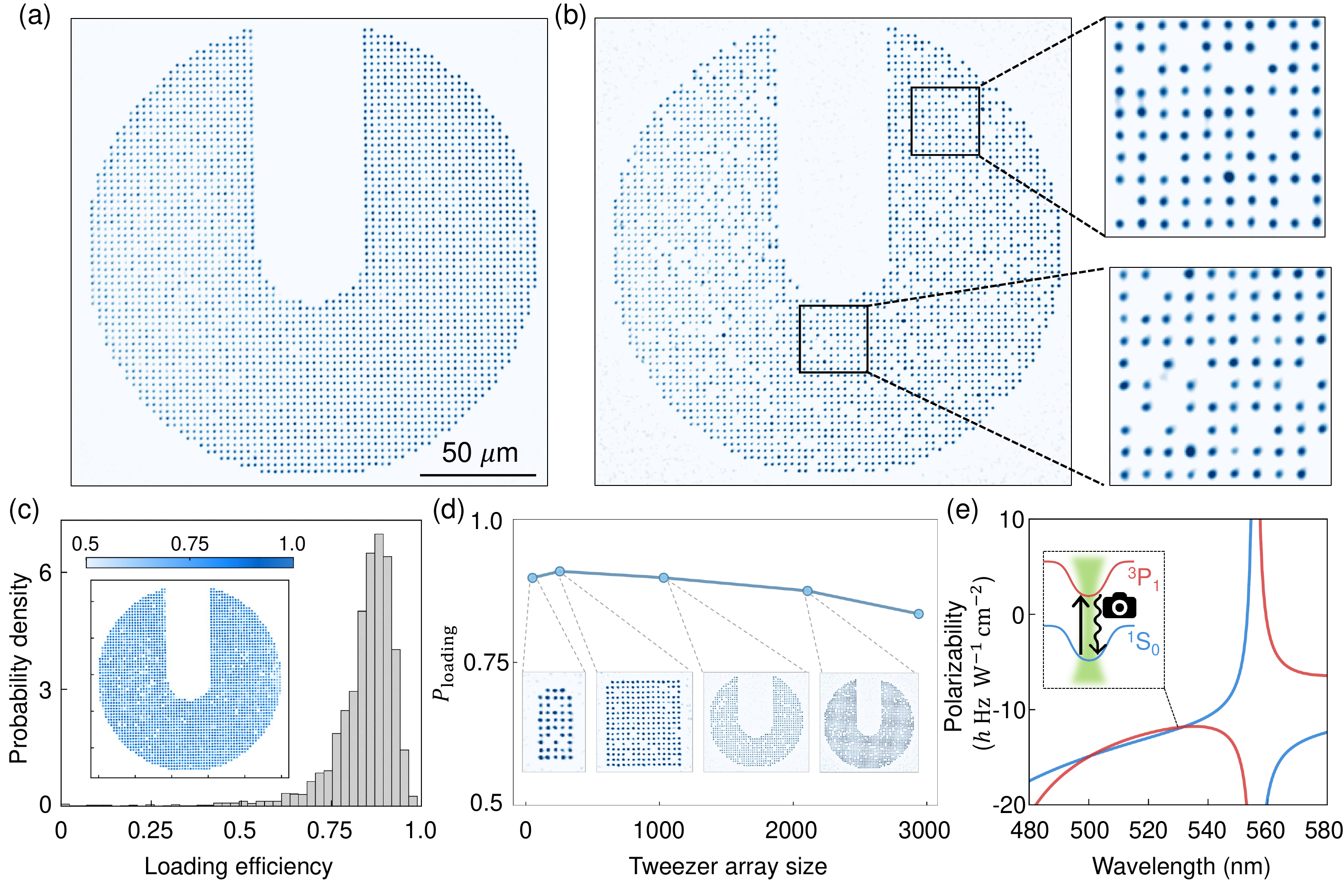}  
\caption{
(a) Averaged fluorescence image of a 2939-site array over 85 experimental trials. 
(b) Single-shot image of 2437 single atoms. Spatial high-pass filter is applied in (a) and (b).
(c) A histogram presenting the site-resolved loading efficiency, with the inset illustrating the efficiency across the array. The average loading efficiency is 83.5(1)\% with standard deviation of 10.3\% across all sites. Data are collected from 85 images.
(d) The loading efficiency for different array sizes with the same trap depth and tweezer spacing. The enhanced loading parameters are optimized for each array size. The highest and lowest loading efficiencies are 91.1(4)\% and 83.5(7)\%, respectively. Error bars are smaller than the marker. Insets show single-shot images of the corresponding array (image of the largest array is in (b)).
(e) Polarizabilities for the ground state $^1S_0$ (blue curve) and the excited state $^{3}P_1$ (red curve) of $^{174}$Yb from numerical calculation (data of atomic levels from Ref.~\cite{energy_level_data_for_polar_cal}), which are used for fluorescence imaging (inset). The optical tweezer operates at the magic wavelength of 532 nm.
The enhanced loading in all plots is performed with $\pi$-polarized light, $I/I_{\text{sat}}=53$, a detuning of $\Delta=2\pi \times4$~MHz relative to the $m_J = 0$ state, and a $2.3\,\text{G}$ magnetic field.
}
\label{fig:1}
\end{figure*}

In this work, we demonstrate a high-efficiency preparation of the largest alkaline-earth-like atom array in optical tweezers, comprising over 2400 ytterbium-174 atoms (Fig.~\ref{fig:1}(a) and 1(b)). 
This is enabled by the efficient trapping and high-fidelity imaging of ytterbium-174 atoms in 532 nm optical tweezers.
We achieve an enhanced loading efficiency of 83.5(1)\% in $\sim 3,000$ optical tweezers (Fig.~\ref{fig:1}(c)).
The high loading efficiency is maintained from a few tens to thousands of tweezers (Fig.~\ref{fig:1}(d)), highlighting the scalability of the enhanced loading method to large arrays.
To understand the light-assisted collision dynamics, we develop a quantitative model that directly links the inelastic collision rate to the observed loading efficiency, yielding good agreement with experiment.
By engineering different interatomic potentials, we demonstrate that enhanced loading does not necessarily require the globally repulsive potentials in earlier experiments~\cite{2015-enhanceloading,2010np-enhanceloading,2019PRX-enhanceloading,kaufman2022PRX-enhanceloading,Aliyu2021prr-enhanceloadingna23,Bryce2022prr-enhanceloadingk39}, but can also be achieved with partially repulsive potentials accessible in generic two-level systems, thereby broadening its applicability across atomic species.
For quantum controls of ytterbium-174 atoms, we propose to encode the qubit in the ground-clock state manifold.
From numerical simulation and comparison to the $^{171}$Yb isotope, we estimate that single-qubit gate fidelity of 99.8\% and two-qubit gate fidelity of 99.9\% can be achieved with experimentally realistic parameters. 
Our work realizes large scale alkaline-earth-like atom arrays with high loading efficiency.
In combination with their estimated long coherence time and high-fidelity controls, alkaline-earth-like atom arrays hold great potential to unlock tremendous capabilities toward universal quantum computation. 

\begin{figure}[t] 
\centering 
\includegraphics[width=0.9\linewidth]{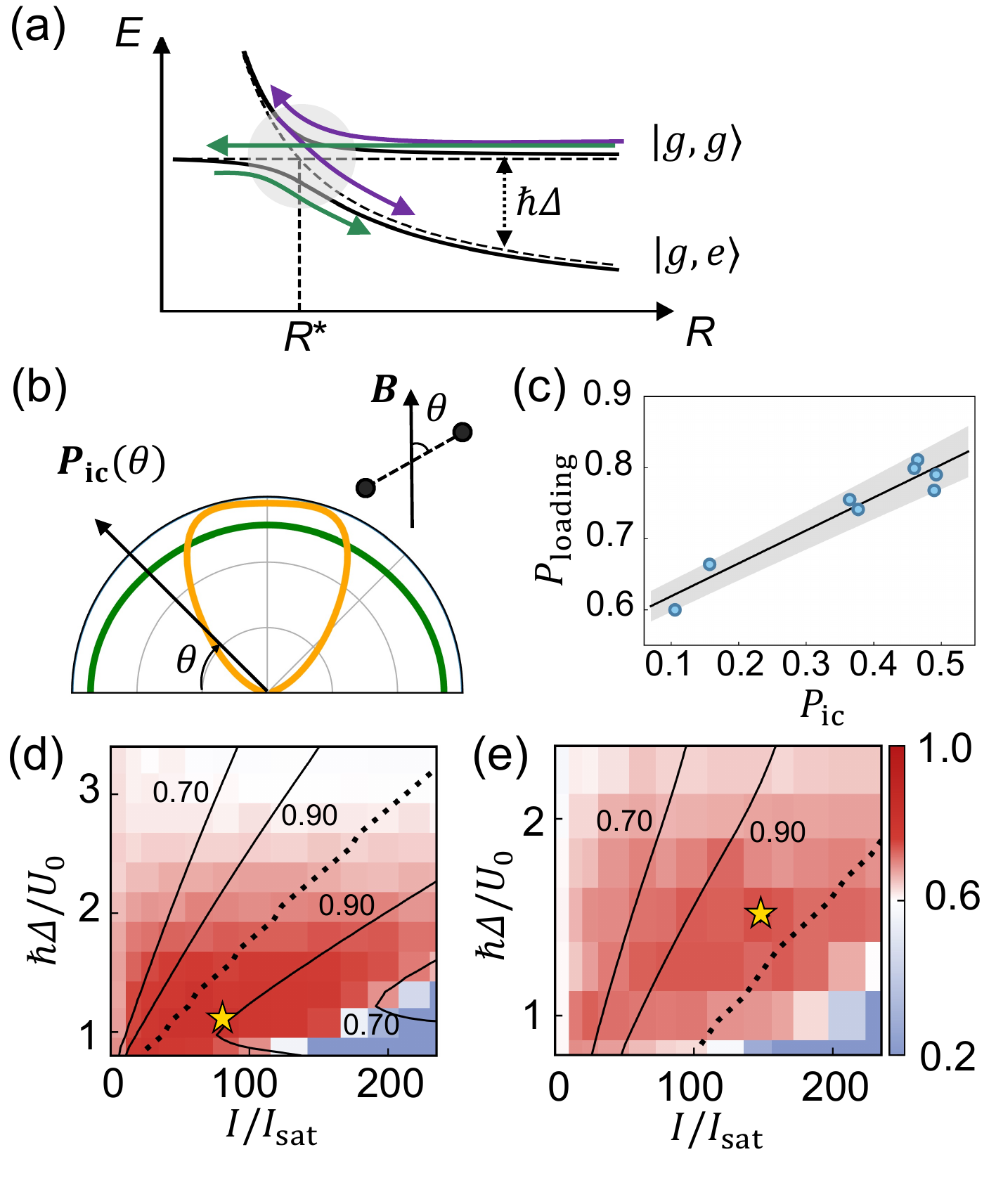}
\caption{
(a) The enhanced loading mechanism. The solid black curves represent the molecular eigenstates in the rotating frame of the blue-detuned laser, with the gray region highlighting the crossing point. Green and purple arrows mark two pathways for inelastic collision.
(b) The angular distribution of the simulated inelastic collision probability $P_\text{ic}$ as a function of $\theta$, with $\theta$ being the angle between the interatomic displacement and the quantization axis ($I/I_{\text{sat}}$ = 100, $\hbar\Delta /U_0$ = 1.35, close to the optimal point in (d), (e)). The green curve corresponds to a globally repulsive potential, while the orange curve corresponds to a potential that is repulsive only around $\theta=90^\circ$.
(c) Linear relationship between the optimal $P_\text{loading}$ and $P_\text{ic}$ for different light polarizations and magnetic fields (data from (d), (e) and Supplemental Material Fig. S3(a),(b)~\cite{SM}). Both $P_\text{loading}$ and $P_\text{ic}$ are maximized over $\hbar\Delta/U_0\in[1, 2]$ and $I/I_\text{sat}\le150$. The black line shows a linear fit, with $\pm1\sigma$ uncertainty marked by the gray shade. The error bars are smaller than the markers.
(d), (e) The loading efficiency of the globally and partially repulsive potential, respectively. The loading enhancement time is 500~ms. The stars label the maximum $P_\text{loading}$ (81.1(4)\% in (d) and 74.1(7)\% in (e)).
The black dots (contours) mark the highest $P_\text{ic}$ (90\% and 70\% of the highest $P_\text{ic}$) for each detuning. 
The detuning is relative to the single-atom energy level coupled by the collision light, i.e. \( m_J = \pm 1 \) in (d) and \( m_J = 0 \) in (e). All data presented in this figure are collected with an array size of 2107.
\label{fig:2}
}
\end{figure} 
\textit{Large-scale $^{174}$Yb atom arrays.---}
Here, we introduce the main results and methods regarding the large-scale atom arrays. We first cool $^{174}$Yb atoms using a two-stage laser cooling setup, resulting in  an ensemble of atoms at the Doppler temperature of 5$\,\mu$K in the magneto-optical trap (MOT).  The detailed implementation is described in Supplemental Material Sec. \MakeUppercase{\romannumeral 1}~\cite{SM}. Static tweezer arrays are generated via a programmable spatial light modulator (SLM) using a weighted Gerchberg-Saxton algorithm \cite{2014PRX-WGS,2019-WGS,Gerchberg1972Optik-WGS}.
A 532 nm laser is phase modulated by the SLM and then focused by a microscope objective (numerical aperture = 0.6, field of view = 500$\,\mu$m), creating arrays of up to 2939 sites. The averaged image is shown in Fig.~\ref{fig:1}(a) and a single-shot image is shown in Fig.~\ref{fig:1}(b). The tweezer spacing is set to 2.8\,$\mu$m and the trap depth is $U_0/\hbar \approx 2\pi\times4$~MHz (3 mW per tweezer, estimated from $^{3}\text{P}_{1}$ tensor light shift).

Cooling and fluorescence imaging of atoms are performed on the $|^{1}\text{S}_{0}\rangle \to|^{3}\text{P}_{1}, m_J=0\rangle$ intercombination line. This transition is magic at the 532 nm tweezer wavelength~\cite{jeff2019PRL-energelevel}, i.e. ac Stark shifts of the ground and excited states are identical (Fig.~\ref{fig:1}(e)). The magic trap enables an imaging survival probability of 99.1(3)\% and an imaging fidelity of 99.3(1)\%.

Leveraging blue-detuned light-assisted collisions, we achieve a high loading efficiency, enabling the capture of more than 2400 atoms in a single experimental trial.  The statistics and spatial distribution of site-resolved loading efficiency are presented in Fig.~\ref{fig:1}(c), showing an average loading efficiency of 83.5(1)\%, where the lower-efficiency sites are mainly attributed to local trap quality that could be mitigated using the method described in Ref.~\cite{endres2025nature-6100atom}. 
We measure loading efficiency for different array sizes and observe over 80\% loading efficiency from tens to thousands of optical tweezers (Fig.~\ref{fig:1}(d)). The moderate decrease for larger arrays is likely due to reduced effective atom density in the target plane \cite{kaufman2022PRX-enhanceloading}, arising from atom capture by residual out-of-plane traps during loading.

\begin{figure}[t]  
\centering 
\includegraphics[width=0.9\linewidth]{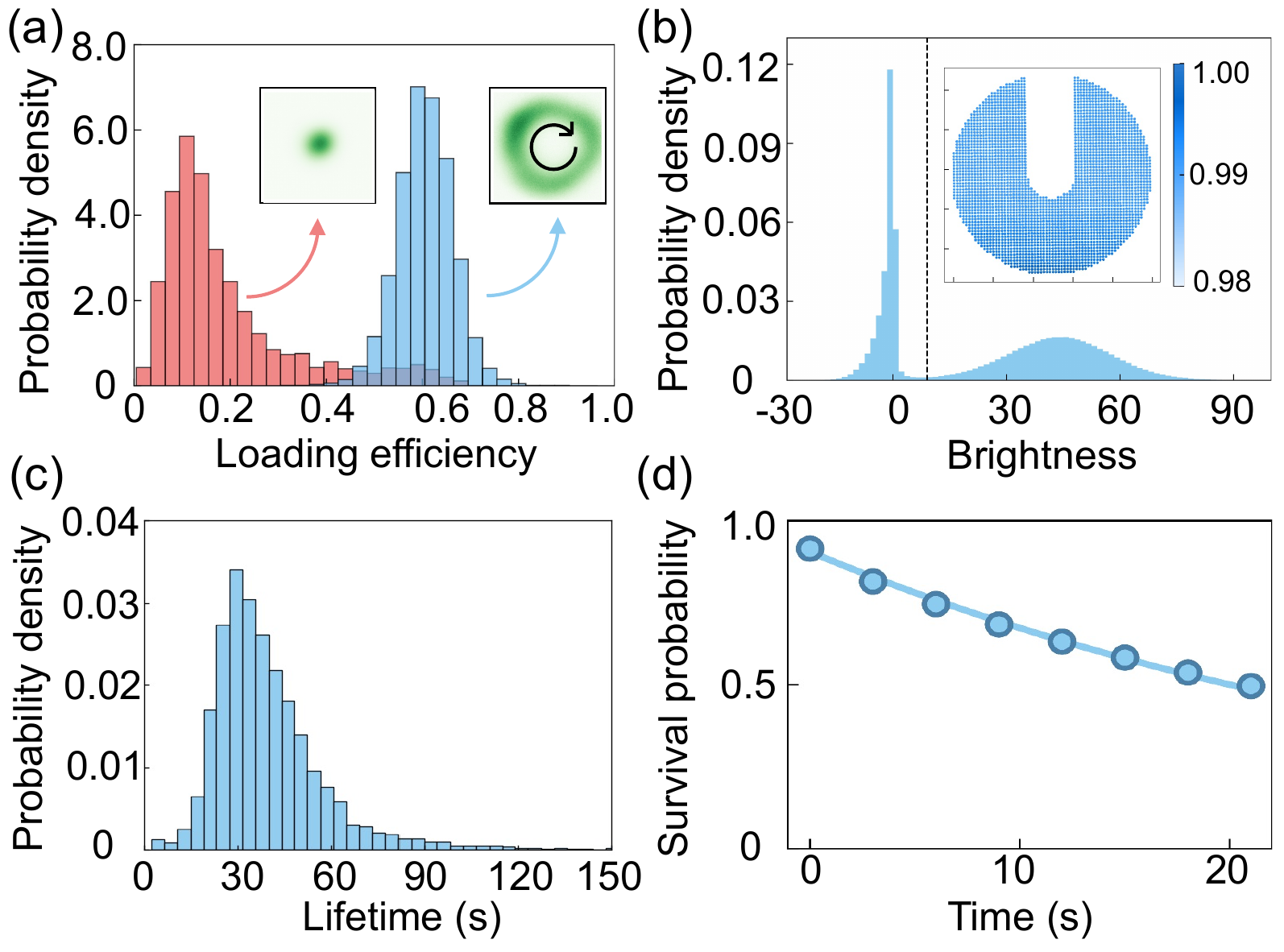} 
\caption{
(a) Initial loading efficiency distribution with (blue) and without (red) implementing the rotating MOT technique, prior to the enhanced loading step. The insets illustrate the fixed MOT and the rotating MOT. 
(b) Histogram of fluorescence brightness across all array sites. The two peaks of the bimodal distribution correspond to zero and one atom occupancy. The inset depicts the site-resolved imaging fidelity. 
The exposure time is 200 ms, limited by the background of out-of-plane atoms (Supplemental Material Sec. \MakeUppercase{\romannumeral 2}~\cite{SM}). Data are collected from 1000 images.
(c) Probability density distribution of imaging lifetime across the array sites. 
(d)The average atom survival over all sites with respect to the imaging time, showing a lifetime of 35.1(5) s.
All data presented in this figure are collected under the maximum array size of 2939.
}
\label{fig:3}
\end{figure} 

\textit{High-efficiency loading.---}
Let us first discuss conventional single-atom preparation methods. A red detuned laser induces light-assisted collisions that expel two atoms out of the tweezer. The combination of the random initial loading number and the two-to-zero collisional loss imposes a $\approx 50\%$ upper bound on the single-atom loading efficiency. 
For atoms with a repulsive interatomic potential, a blue-detuned collision laser has been used to improve the loading efficiency in certain cases~\cite{2015-enhanceloading,2010np-enhanceloading,2019PRX-enhanceloading,kaufman2022PRX-enhanceloading,Aliyu2021prr-enhanceloadingna23,Bryce2022prr-enhanceloadingk39}. However, it remains elusive whether this method can be applied to other atomic species and what the optimal parameters for loading enhancement are.

Here we present a quantitative model of the blue-detuned light-assisted collision process that answers the above questions. Figure~\ref{fig:2}(a) shows the interatomic potential in the rotating frame of the collision laser frequency, considering one ground state $|g\rangle$ and one excited state $|e\rangle$ for simplicity. During the collision process, an atom pair starting from the $|g,g\rangle$ state passes through the $|g,g\rangle$–$|g,e\rangle$ crossing twice. If \textit{one of the two passages is adiabatic and the other is nonadiabatic} (green and purple arrows in Fig.~\ref{fig:2}(a)), an inelastic collision happens and the atom pair gains kinetic energy $\hbar\Delta$, with $\Delta$ being the laser detuning. When $\hbar\Delta \gtrsim U_{\text{trap}}$, only one atom can escape from the trap, leading to a one-by-one atom loss until only a single atom remains~\cite{Scazza2025PRL—bluedetuned}. 
We calculate the interatomic potential using the dipole-dipole interaction model and the inelastic collision probability $P_\text{ic}$ by solving the Lindblad equation~(Supplemental Material  Sec. \MakeUppercase{\romannumeral 3}~\cite{SM}). 

Next, we experimentally study the enhanced loading performance by engineering interatomic potentials. 
We realize two representative potentials -- (i) a globally repulsive potential using mixed light polarizations under zero magnetic field (the green curve in Fig. \ref{fig:2}(b)) and (ii) a partially repulsive potential using pure $\pi$ polarization under 8.6~G field along the tweezer polarization direction~(the orange curve in Fig. \ref{fig:2}(b)). 
We observe significant loading enhancement under both the globally and partially repulsive potentials (Fig.~\ref{fig:2}(d),(e)). We also find good agreement between the calculated $P_\text{ic}$ and the measured loading efficiency, validating the theoretical model (Fig.~\ref{fig:2}(c)-(e)). The maximum loading efficiency achieved here is $\sim80\%$. The imperfection is because we apply a cooling light during collision (red detuned from the $|^1\text{S}_0\rangle\to|^3\text{P}_1, m_J=0\rangle$ magic transition) to cool the remaining atom, which induces a finite two-to-zero loss probability as in the conventional red-detuned light-assisted collision process. Enhanced loading performance under other potentials can be found in  Supplemental Material Fig. S3~\cite{SM}. 

Note that previous enhanced loading studies~\cite{2015-enhanceloading,2010np-enhanceloading,2019PRX-enhanceloading,kaufman2022PRX-enhanceloading,Aliyu2021prr-enhanceloadingna23,Bryce2022prr-enhanceloadingk39} utilize either $\Lambda$ systems or multiple degenerate states where the interatomic potentials are globally repulsive, similar to the case in Fig.~\ref{fig:2}(d). Here, we find loading enhancement under partially repulsive potentials, which are attainable in generic two-level systems (Supplemental Material Sec. \MakeUppercase{\romannumeral 3}~\cite{SM}). 

\textit{Characterization of atom loading and imaging.---}
As the tweezer array size (82 $\mu$m radius) is larger than the MOT ($1/e^2$ radius of 48 $\mu$m), we rotate the position of the MOT to achieve a uniform loading by modulating the bias magnetic field. Figure~\ref{fig:3}(a) shows a significant improvement of initial loading efficiency with MOT rotation. More details about the rotating MOT method can be found in Supplemental Material Sec. \MakeUppercase{\romannumeral 1}~\cite{SM}. 

Next, we characterize the imaging fidelity. 
As the large optical tweezer array is inevitably associated with out-of-plane traps~\cite{endres2025nature-6100atom}, fluorescence from atoms in those traps creates a strong background in the image. We apply spatial high-pass filters to eliminate this background following Ref.~\cite{ustc2025PRL-2024atom} (Supplemental Material Sec. \MakeUppercase{\romannumeral 2}~\cite{SM}).
Figure~\ref{fig:3}(b) displays the statistics of fluorescence brightness for all sites in 1,000 images. 
We use the same brightness threshold to determine the occupancy for all sites. 
Using the analysis method from Ref.~\cite{endres2025nature-6100atom}, we estimate the average imaging fidelity of 99.3(1)\% and the atom loss rate of 0.9(3)\% per image. 
The spatial distribution of imaging fidelities across the entire array is presented in the inset in Fig.~\ref{fig:3}(b).  
Imaging lifetime is determined by continuously shining the imaging light and measuring atom survival probabilities. The result is presented in Fig.~\ref{fig:3}(c),(d) with an average imaging lifetime of 35.1(5) s, consistent with the atom loss rate measured above.


\begin{figure}[t]  
\centering 
\includegraphics[width=0.9\linewidth]{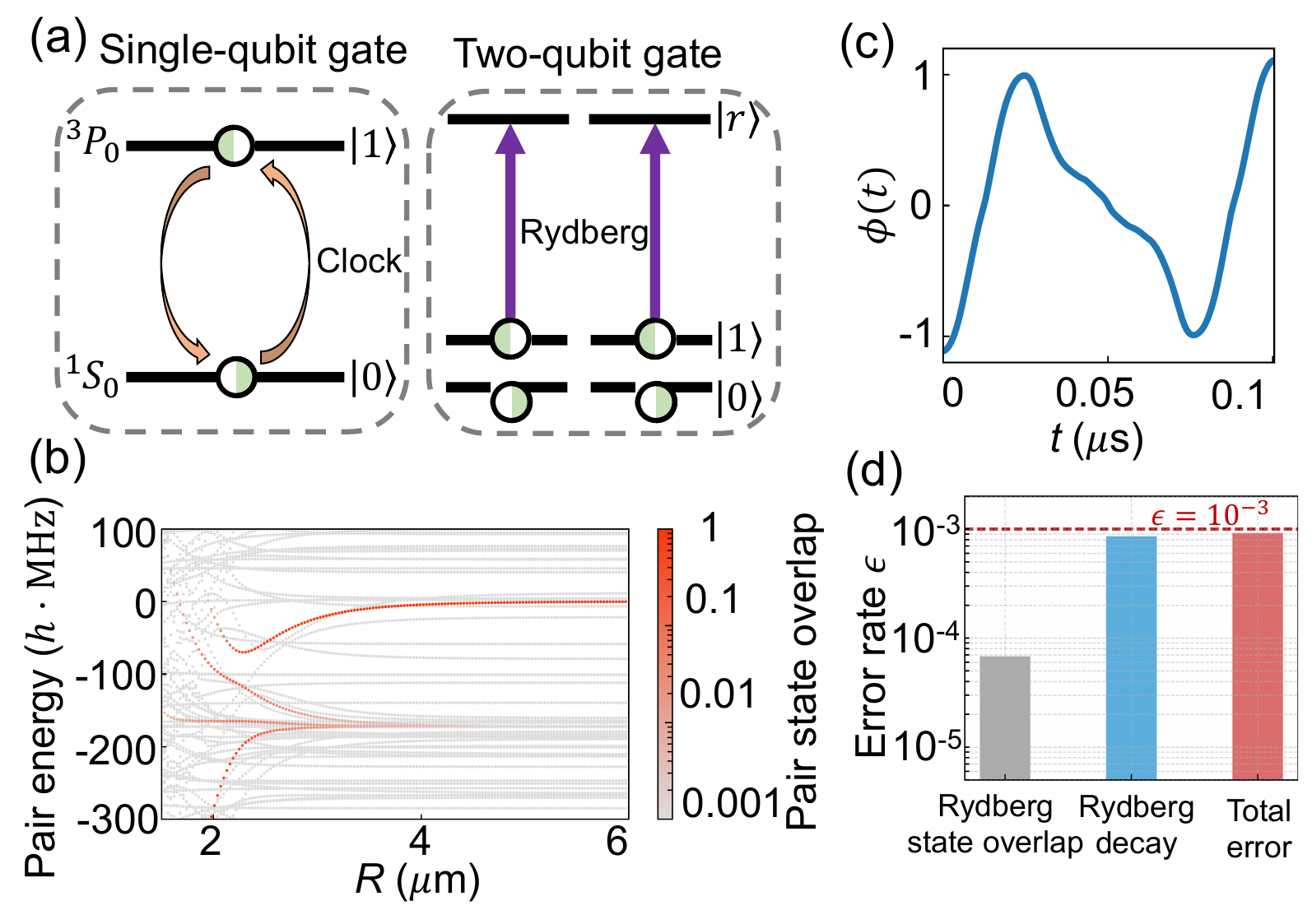}  
\caption{
(a) Level diagram of single- and two-qubit gates.
(b) Calculated Rydberg-pair interaction potential for the state $|r\rangle = |\nu=49.56, L=0, J=1, m_J=1\rangle$ at a magnetic field of 200 G with atoms aligned perpendicular to the magnetic field. The color of each curve represents its overlap with the target pair state $|r,r\rangle$. 
(c) Laser phase of the time-optimal CZ gate, assuming an interatomic distance of 1.9 $\mu$m.
(d) Numerical simulation of the two-qubit gate error rates.}
\label{fig4}
\end{figure}
\textit{$^{174}$Yb gate scheme.---}
To establish the large-scale $^{174}$Yb atom arrays as quantum processors, we next present qubit encoding and quantum gate schemes. Compared to the commonly used $^{171}$Yb isotope with well-studied nuclear spin qubits~\cite{jeff2023nature-erasure, kaufman2023PRX-omg}, $^{174}$Yb has zero nuclear spin and thus lacks the nuclear spin qubits. Inspired by the \textit{omg} encoding demonstrated in $^{171}$Yb~\cite{kaufman2023PRX-omg}, we propose to encode qubits in the ground $^1\text{S}_0$ and clock $^3\text{P}_0$ states of $^{174}$Yb atoms, as shown in Fig.~\ref{fig4}(a).  One advantage of this optical qubit is that $J=0$ for both states, and therefore it is insensitive to magnetic field noise and tweezer polarization imperfections~\cite{Jeff2025arXiv-leaverageErasureError}. The 578 nm clock transition between qubit states can be excited with a magnetic field that mixes $^3P_0$ and $^3P_1$~\cite{barber2006direct}. For a 200 G magnetic field, 1 W laser power (2 W 578 nm laser is commercially available) with a beam waist of 9 $\mu$m can induce a Rabi frequency of $2\pi \times110$~kHz. This is comparable to the 99.8\% single-qubit gate in the $^{171}$Yb case~\cite{kaufman2023PRX-omg}. For the clock-state encoding, the corresponding magic trapping wavelength is 759 nm, where the polarizability and available laser power are smaller than those at 532 nm.  Nevertheless, the available 759 nm optical power is not expected to be the limitation at the demonstrated array scale, because clock-state storage can be achieved with much shallower traps (Supplemental Material Sec. \MakeUppercase{\romannumeral 5}~\cite{SM}).

For two-qubit gates, we propose to couple $^3P_0$ to $|r\rangle=|\nu=49.56,L=0,J=1,m_J=1\rangle$ Rydberg state, where $\nu, L, J, m_J$ are the effective principal quantum number, orbital angular momentum, total angular momentum and total angular momentum z component, respectively. This Rydberg excitation requires a 302 nm UV laser, where $\sim 1$ W power is commercially available. Assuming a 75 mW Rydberg laser inside the vacuum chamber and a beam waist of 12$\,\mu$m, we estimate the Rabi frequency to be $2\pi\times15$~MHz. 
We consider the fundamental errors of two-qubit Rydberg gates -- Rydberg decay and excitation to nearby overlapping Rydberg pair states (Supplemental Material Sec. \MakeUppercase{\romannumeral 5}~\cite{SM}).
We calculate the Rydberg pair state overlaps following the multichannel quantum defect model from Ref.~\cite{jeff2025PRX-gatefidelity} (Fig.~\ref{fig4}(b)). 
With the Rydberg state information, we then optimize a pulse for the controlled-Z (CZ) gate using the gradient ascent pulse engineering algorithm ~\cite{Jeff2023PRX-rydberggate}. The resulting pulse is shown in Fig.~\ref{fig4}(c). The simulated CZ gate fidelity is 99.91\% (quantified using the Bell state fidelity), dominated by Rydberg decay. 
The results demonstrate that the gate fidelities of $^{174}$Yb are well above the surface‑code error‑correction thresholds and are comparable to those of the commonly used $^{171}$Yb isotope. 

We note that the $^{174}$Yb atom array is also promising for quantum simulations with the qubits encoded in the clock-Rydberg manifold \cite{Endres2023Nature-erasure}. Featuring large system sizes and high-fidelity Rydberg excitations, the $^{174}$Yb atom array can simulate quantum many-body systems approaching the thermodynamic limit with high precision.

\textit{Conclusion and outlook.---}
In summary, we realize alkaline-earth-like atom arrays comprising over 2400 ytterbium-174 atoms in optical tweezer arrays. We systematically investigate the enhanced loading mechanism for a range of array sizes and interatomic potentials, demonstrating its scalability to larger arrays and applicability to most atoms or molecules with dipole-dipole interactions~\cite{Doyle2019Science-tweezer,KangKuen2022IOP-tweezer}. We also propose a qubit encoding scheme using the ground and clock states of $^{174}$Yb, and estimate gate fidelities approaching 99.9\%. 

Further improvements of the array size can be achieved by reusing laser power via optical lattices or optical cavity arrays~\cite{Bloch24prl-10000, Simon2025arXiv-cavityarraymicroscopeparallel}. 
We also note that the light-assisted collisions between $J=0$ and $J=1$ states in this work is reminiscent of the microwave shielding of ultracold polar molecules in the lowest rotational states with angular momenta 0 and 1~\cite{Jeremy2018PRL-shield,Goulven2018PRL-sheild,Doyle2021Science-shield,xinyu2022Nature-shield,wangdajun2023PRX-shield,Sebastian2023NP-shield,shitao2023PRL-shield}.
Therefore, our work might provide mutual reference and promotion between the single-atom preparation in atom arrays and microwave shielding of ultracold molecules. 
Finally, the integration of loading enhancement, rapid SLM rearrangement~\cite{ustc2025PRL-2024atom} and continuous reloading techniques~\cite{lukin2025nature-3000atom, li2025fast, MPQ2024PRR-cloading,2024PRX-1225atom} would enable large-scale, defect-free and sustainable $^{174}$Yb atom arrays. The combination of the large qubit number and estimated high gate fidelities makes the $^{174}$Yb arrays a promising platform for quantum error correction and logical operations at large scale~\cite{lukin2025nature-universal}.

\vspace{.5cm}
\textit{Acknowledgments.---}
We thank Wenjun Zhang for fruitful discussions, Rongpei Zhu for the help with the laser setups, and Hao Zhang and Yue Jiang for the help with objective tests. This work was supported by the Quantum Science and Technology–National Science and Technology Major Project under Grant No. 2024ZD0301600 and Grant No. 2025ZD0300400, the National Natural Science Foundation of China under Grants No. 12474479 and No. 12504573, the Beijing Natural Science Foundation under Grant No. F251004, the Innovation Program for Quantum Science and Technology under Grant No. 2023ZD0300700, the National Key Research and Development Program of China under Grant No. 2022YFA1405300, the China Postdoctoral Science Foundation under Grants No. 2024M760064 and No. 2025T180934, and the Postdoctoral Fellowship Program of CPSF under Grant No. GZB20250790.

\textit{Data availability.---}
The data that support the findings of this study are available ~\cite{data}.

\nocite{Preskill2002AIP-quantummemory,Andrew2021PRA-surfacecode,Zhang2024Optica-longcoherencetime,Barnes2022nc-Yblongcoherencetime,}

\bibliography{main}

\end{document}


\title{Supplemental Material for: 

High-efficiency loading of 2,400 Ytterbium atoms in optical tweezer arrays}

\author{Jiawen Zhu}
\altaffiliation{These authors contributed equally to this work.}
\affiliation{State Key Laboratory for Artificial Microstructure and Mesoscopic Physics and Frontiers Science Center for Nano-optoelectronics, School of Physics, Peking University, Beijing 100871, China}

\author{Changfeng Chen}
\altaffiliation{These authors contributed equally to this work.}
\affiliation{State Key Laboratory for Artificial Microstructure and Mesoscopic Physics and Frontiers Science Center for Nano-optoelectronics, School of Physics, Peking University, Beijing 100871, China}

\author{Li Zhou}
\affiliation{State Key Laboratory for Artificial Microstructure and Mesoscopic Physics and Frontiers Science Center for Nano-optoelectronics, School of Physics, Peking University, Beijing 100871, China}

\author{Xiangru Xie}
\affiliation{State Key Laboratory for Artificial Microstructure and Mesoscopic Physics and Frontiers Science Center for Nano-optoelectronics, School of Physics, Peking University, Beijing 100871, China}

\author{Chenyang Jiang}
\affiliation{State Key Laboratory for Artificial Microstructure and Mesoscopic Physics and Frontiers Science Center for Nano-optoelectronics, School of Physics, Peking University, Beijing 100871, China}

\author{Zhuoli Ding}
\affiliation{State Key Laboratory for Artificial Microstructure and Mesoscopic Physics and Frontiers Science Center for Nano-optoelectronics, School of Physics, Peking University, Beijing 100871, China}

\author{Fan Wu}
\affiliation{State Key Laboratory for Artificial Microstructure and Mesoscopic Physics and Frontiers Science Center for Nano-optoelectronics, School of Physics, Peking University, Beijing 100871, China}

\author{Fan Yang}
\affiliation{Hefei National Laboratory, Hefei 230088, China}

\author{Guoqing Wang}
\affiliation{International Center for Quantum Materials, School of Physics, Peking University, Beijing 100871, China}

\author{Qihuang Gong}
\affiliation{State Key Laboratory for Artificial Microstructure and Mesoscopic Physics and Frontiers Science Center for Nano-optoelectronics, School of Physics, Peking University, Beijing 100871, China}
\affiliation{Hefei National Laboratory, Hefei 230088, China}
\affiliation{Collaborative Innovation Center of Extreme Optics, Shanxi University, Taiyuan 030006, China}
\affiliation{Peking University Yangtze Delta Institute of Optoelectronics, Nantong, Jiangsu 226010, China}

\author{Peng Zhang}
\affiliation{School of Physics, Renmin University of China, Beijing 100872, China}
\affiliation{Key Laboratory of Quantum State Construction and Manipulation (Ministry of Education), Renmin University of China, Beijing 100872, China}

\author{Sheng Zhang}
\email{sheng.physik@pku.edu.cn}
\affiliation{State Key Laboratory for Artificial Microstructure and Mesoscopic Physics and Frontiers Science Center for Nano-optoelectronics, School of Physics, Peking University, Beijing 100871, China}

\author{Pai Peng}
\email{pengpai@pku.edu.cn}
\affiliation{State Key Laboratory for Artificial Microstructure and Mesoscopic Physics and Frontiers Science Center for Nano-optoelectronics, School of Physics, Peking University, Beijing 100871, China}
\affiliation{Collaborative Innovation Center of Extreme Optics, Shanxi University, Taiyuan 030006, China}
\maketitle
\section{Apparatus and Methods for Large-Scale Atom Arrays}
\begin{figure}
\centering
\includegraphics[width=0.5\linewidth]{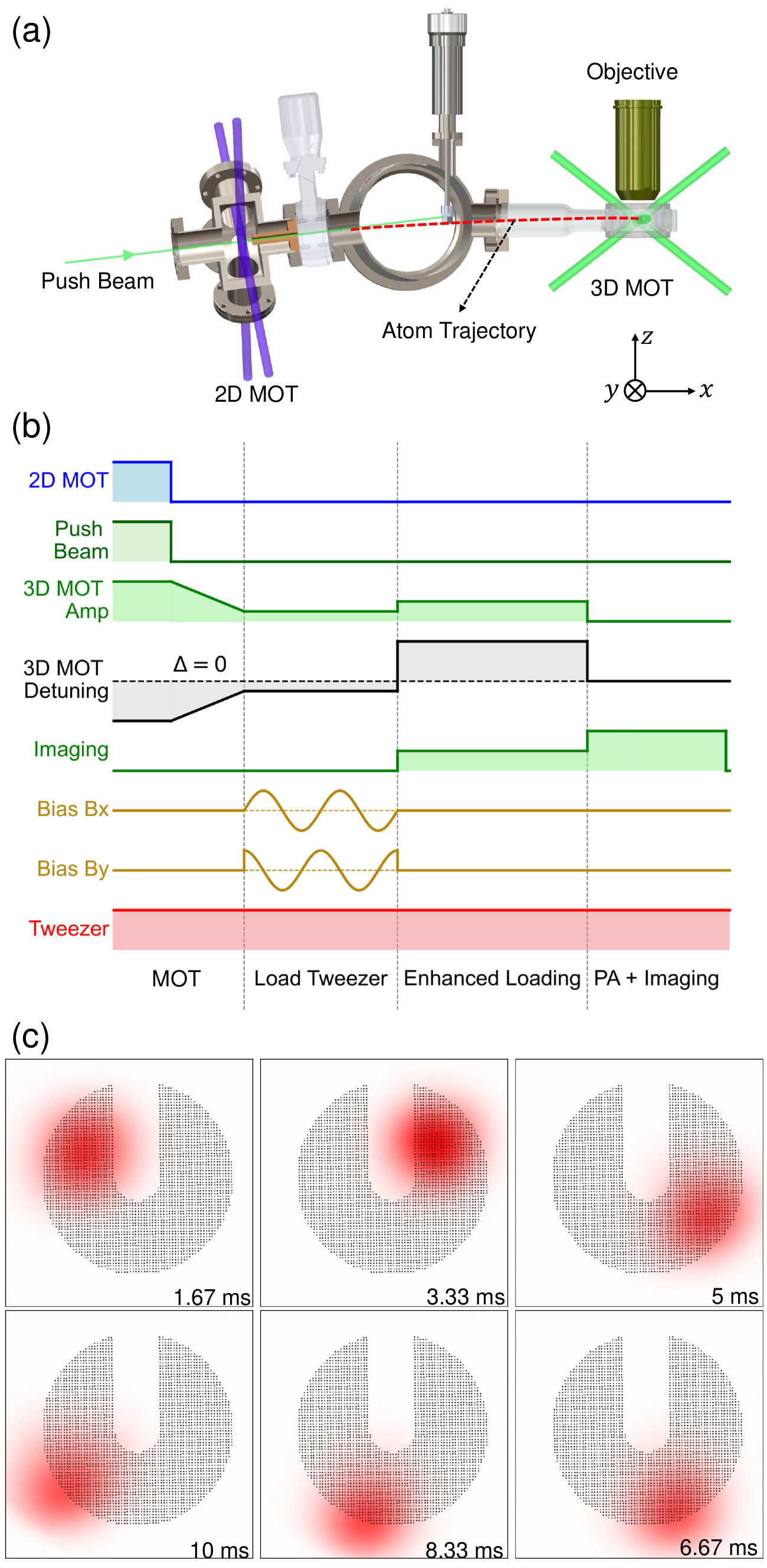}
\caption{(a) Schematic of the experimental apparatus. Purple beams denote the 399 nm cooling light used for the 2D MOT. A weak 556 nm beam (green) entering from the left acts as a push beam, transporting atoms into the science chamber before being blocked by an aluminum mirror. The red dashed line indicates the resulting atomic trajectory. Within the science chamber, three orthogonal 556 nm beams form a 3D MOT. An objective mounted above the chamber focuses the 532 nm laser to generate optical tweezers and simultaneously collects single-atom fluorescence for detection.
(b) Time Sequence for MOT loading, tweezer loading, enhanced loading and imaging.   
(c) The MOT positions at different moments during the loading process. The red shade represents fluorescence of atoms in the MOT, and the black dotted grid indicates the positions of the optical tweezer arrays.
    }
\label{fig:setup}
\end{figure}

The schematic diagram of the vacuum system is shown in Fig.~\ref{fig:setup}(a). To prepare a large ensemble of cold $^{174}\mathrm{Yb}$ atoms, we employ a two-stage laser cooling scheme. Atoms emitted from a thermal oven are first captured and pre-cooled in a two-dimensional magneto-optical trap (2D MOT) operating on the broad $^{1}\text{S}_0\to\,^{1}\text{P}_1$ transition at 399 nm.  The quadrupole magnetic field of the 2D MOT is generated by eight permanent magnets, producing a field gradient of 49~G/cm. The pre-cooled atoms are subsequently transported into the science chamber, where they are captured by a three-dimensional MOT (3D MOT) operating on the narrow $^{1}\text{S}_0\to\,^{3}\text{P}_1$ intercombinaion transition at 556 nm. The 2D MOT chamber and the science chamber are separated by a differential pumping tube, which maintains a pressure difference exceeding one order of magnitude.

\label{cooling}
Atoms are transported from the 2D MOT to the 3D MOT using a 556~nm push beam, following the approach of Ref.~\cite{jeff2019PRL-energelevel}. The push beam is aligned along the system axis to push the atoms out of the 2D MOT. After passing through the differential pumping tube, the atomic cloud separates from the push beam and follows a parabolic trajectory into the 3D MOT under gravity. To obtain a high number of trapped atoms, the 3D MOT initially operates with high laser power and large detuning, and an electro-optic modulator is used to generate frequency sidebands on the MOT beams. After loading, the modulator is turned off and the MOT power and detuning are ramped down within 10~ms, further cooling the atoms to a temperature close to the Doppler limit of approximately 5~$\mu$K.

A high-power 532~nm laser (40~W, Precilaser) is used to generate a static optical tweezer array. The beam is phase-modulated by a spatial light modulator (UPOLabs HDSLM80R Plus) with a resolution of $1920\times1200$ and a refresh rate of 60~Hz, and then focused into the science chamber by a custom-designed objective lens (Special Optics) with a numerical aperture of 0.6 and a field of view of 500~$\mu$m. Here we use only a small region (< 200 $\mu$m) of the field of view to minimize the diffraction angle and thus maximize diffraction efficiency of the SLM. Fluorescence photons are collected by the same objective and detected by a low-noise CMOS camera (Hamamatsu ORCA-QUEST C15550-20UP). The strong zeroth-order diffraction light from the SLM impairs the MOT and creates unwanted out-of-plane traps. We block it using a bar-shaped mirror which also blocks the region above the array center, as shown in Fig.~\ref{fig:1}(a),(b) in the main text. 


In our current implementation, the most immediate limitation is the optical power delivered to the tweezer array, set by the available 532 nm laser power and the SLM diffraction efficiency. We use a 40 W 532 nm laser and a broadband SLM with an efficiency of about 25\%, which supports approximately 3,000 tweezers at our typical loading/imaging power of 3 mW per tweezer. Using state-of-the-art 532 nm CW lasers with up to 80 W power and reported SLM diffraction efficiencies of about 37\% for 10,000-site arrays~\cite{jeff2024optica-localgate}, the delivered power would be about 30 W, corresponding to approximately 10,000 tweezers at comparable trap depth.

Other factors do not appear to impose a comparable near-term limit. A larger objective field of view, such as the 1.5 mm FOV used in Ref.~\cite{endres2025nature-6100atom}, would accommodate arrays well beyond 10,000 sites at similar spacing. Trap-depth uniformity can be improved with atom-based feedback, as demonstrated in Ref.~\cite{endres2025nature-6100atom} for a 12,000-site array. The rotating MOT technique reduces constraints from the MOT size, and the atom flux can be further increased, for example by adding a Zeeman slower. Imaging background from out-of-plane traps can be suppressed by selective removal~\cite{endres2025nature-6100atom}, although its ultimate impact at larger scales requires further study. Finally, current high-resolution SLMs provide sufficient optical degrees of freedom for arrays beyond 10,000 sites~\cite{jeff2024optica-localgate}. These estimates indicate that scaling to the 10,000 tweezer level is technically feasible.

Because the tweezer array spans a larger area than the MOT, the MOT is dynamically moved to achieve homogeneous loading across the array (Fig.~\ref{fig:3}(a) in the main text). As illustrated in  Fig.~\ref{fig:setup}(b), ac currents with a relative phase of $\pi/2$ are applied to the compensation coils along the $x$ and $y$ directions, generating a rotating bias magnetic field that drives the MOT to rotate around the tweezer sites. During the 100~ms tweezer loading stage, the bias field is modulated with a 10~ms period. Figure~\ref{fig:setup}(c) shows the MOT positions at different moments. Enhanced loading is then realized by applying blue-detuned collision light together with cooling light for 500~ms, after which a 100~ms pulse of high-power imaging light induces red-detuned light-assisted collisions (photon association, PA) to eliminate multiply occupied sites. Figure~\ref{fig:setup}(b) shows one of the laser beam schemes used to enhance the loading efficiency. Here, the MOT beam (mixed polarization) is used as blue-detuned collision light. The $\pi$-polarized imaging beam is used for cooling, with $I/I_{\text{sat}} = 15$ and $\Delta = -2\pi \times 0.26$ MHz relative to the $m_J = 0$ transition. In other cases (e.g. Fig.~\ref{fig:2}(e) in the main text), we use the imaging beam as blue-detuned collision light and the MOT beam as cooling light. 
\section{Image analysis methods}
\label{app:image}
To determine whether each site is occupied, we select a 10×10 pixel region of interest (ROI) for each site and calculate a weighted summation of the fluorescence counts within that ROI. The weight is proportional to the average brightness profile of atoms. Specifically, we continuously capture 50 images and use the normalized average brightness as the weight. The brightness histogram for each site exhibits a bimodal distribution, with two peaks corresponding to background (atom absent) and single-atom fluorescence (atom present), respectively, as shown in Fig.~\ref{fig:3}(b) in the main text. 
\begin{figure}
    \centering
    \includegraphics[width=0.5\linewidth]{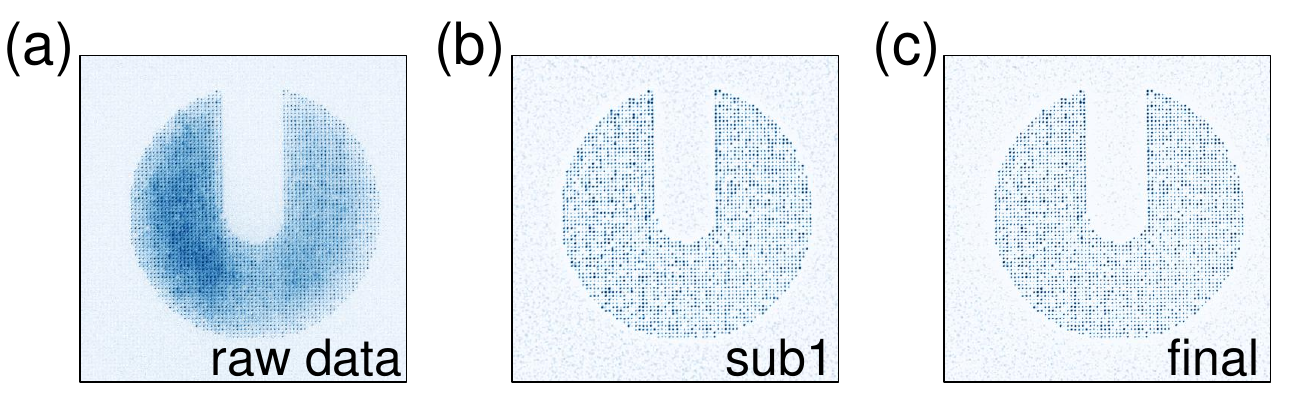}
    \caption{(a) Raw data $I_{\mathrm{raw}}$. (b) Intermediate image after the first background subtraction $I_{\mathrm{sub1}}$. (c) Final processed image $I_{\mathrm{final}}$.}
    \label{filter}
\end{figure}
A key challenge in performing high-fidelity imaging in the large-scale atom arrays is the added background noise from out-of-plane atoms. The interference of multiple tweezer beams generates numerous unwanted out-of-plane traps, where a large number of atoms are trapped~\cite{endres2025nature-6100atom, ustc2025PRL-2024atom}. As shown in~ Fig.~\ref{filter}(a), this background noise severely degrades the signal-to-noise ratio of the fluorescence from the in-plane atom array, making it difficult to achieve high-fidelity detection of atoms. As the fluorescence from out-of-plane atoms are defocused at the camera, adverse effect can be mitigated using spatial high-pass filtering~\cite{ustc2025PRL-2024atom}. We apply three Gaussian filters to the raw data: the first filter uses a sharp kernel to preserve the fine structure of single-atom signals; the second employs a medium‑sized kernel for an initial suppression of background noise; and the third utilizes a larger kernel to reduce large‑scale background fluctuations. Specifically, the images are processed as follows:

\begin{equation}
\begin{aligned}
I_{\text{sub1}} &= \max\left\{ I_{\text{raw}} * G(\sigma_{\text{sharp}}) - I_{\text{raw}} * G(\sigma_{\text{wide1}}), 0 \right\}, \\
I_{\text{final}} &= \max\left\{ I_{\text{sub1}} - I_{\text{raw}} * G(\sigma_{\text{wide2}}), 0 \right\},
\end{aligned}
\end{equation}
where $G(\sigma)$ denotes a two‑dimensional Gaussian kernel with standard deviation $\sigma$, $*$ represents the two‑dimensional convolution operation, and the function $\max\{\cdot,0\}$ truncates the result at zero to eliminate negative pixel values introduced by background subtraction. Here, $I_{\text{raw}}$ is the original fluorescence image acquired by the camera, $I_{\text{sub1}}$ is the intermediate image obtained after the first stage of background removal, and $I_{\text{final}}$ is the high‑contrast image output after the second filtering stage. The kernel parameters $(\sigma_{\text{sharp}}, \sigma_{\text{wide1}}, \sigma_{\text{wide2}}) = (2.4, 16.8, 91.9)$ are optimized by minimizing the overlap between the two peaks in the brightness histogram of the filtered image data.

The effect of the filtering process can be seen in Fig.~\ref{filter}. 
Although the fluorescence of out-of-plane atoms are significantly reduced by filtering, it is still the major source of the image background and limitation for the exposure time. This could be mitigated by blowing out the out-of-plane atoms using laser pulses as shown in Ref.~\cite{endres2025nature-6100atom}. Alternatively, we expect that the subsequent rearrangement process would also eliminate most of the out-of-plane atoms without any additional treatments.
 
After filtering the out-of-plane atoms, we adopt the threshold and calibration method in Ref.~\cite{endres2025nature-6100atom}. 
Leveraging the temporal correlation of three consecutive images, we determine both the atom loss rate and the false positive/negative classification errors. We optimize the threshold value with respect to the imaging fidelity.

\vspace{.5cm}
\section{Theoretical models of blue-detuned light-assisted collision}
\label{theory}
\subsubsection{Hamiltonian}
Let us consider a four-level system as in our experiment, consists of one ground state $\ket{g} = \ket{^1S_0}$ and three excited states $\ket{e_j} = \ket{^3P_1, m_J = j}, \text{where}\, j\in\{-1, 0, 1\}$. The single-atom Hamiltonian in the rotating frame of the blue-detuned collision light is given by
\begin{equation}
    H_0 = \hbar\Delta\ket{g}\bra{g} + \sum_j\big(\hbar\frac{\Omega_j}{2}\ket{g}\bra{e_j}+{\rm H.c.}\big),
\end{equation}
where the rotating-wave approximation (RWA) has been applied.
The tensor light shift and zeeman shift is given by
\begin{equation}
    \begin{aligned}
    H_{\text{shift}} =& \hbar\Delta_{\text{TLS}}\big(\ket{e_1}\bra{e_1}+\ket{e_{-1}}\bra{e_{-1}}\big) +\\
    & \hbar\Delta_{\text{Zeeman}}\big(\ket{e_1}\bra{e_1}-\ket{e_{-1}}\bra{e_{-1}}\big).
    \end{aligned}
\end{equation}
The dipole-dipole interaction between two atoms is (in the Schr{\"o}dinger picture)
\begin{equation}
  \hat{V}_{\rm dd}({\bf r}) = \frac{1}{4\pi\epsilon_0R^3}\left[\hat{\bf d}^{(1)}\cdot\hat{\bf d}^{(2)}-3(\hat{\bf d}^{(1)}\cdot\hat{\bf r})(\hat{\bf d}^{(2)}\cdot\hat{\bf r})\right],
\end{equation}
with $\bf R$ being the interatomic displacement, $\bf d^{(i)}$ being the dipole operator of the $i$-th atom, with $i=1,2$.

Under the RWA, only the energy-conserving terms in $\hat{V}_{\text{dd}}$ are of interest. Moreover, since $1\leftrightarrow 2$ exchange symmetry of $\hat V_{\rm dd}$ ensures parity conservation and the initial state $|g,g\rangle$ is symmetric, we restrict our analysis to the even-parity sector. Define symmetric state $\ket{ge_j}_\text{s}\equiv \left(\ket{g}_1\ket{e_j}_2+\ket{e_j}_1\ket{g}_2\right)/\sqrt{2}$. In the basis of $\{\ket{ge_{-1}}_\text{s}, \ket{ge_{0}}_\text{s}, \ket{ge_1}_\text{s} \}$, $\hat{V}_{\text{dd}}$ can be written as
\[
\begin{gathered}
\hat{V}_{\text{dd}} = \frac{d^2}{4\pi\epsilon_0R^3}\times\\
\begin{pmatrix}\dfrac{3\cos^2\theta-1}{2} & -\dfrac{3\sin2\theta e^{-i\phi}}{2\sqrt2} & -\dfrac{3\sin^2\theta e^{-i2\phi}}{2}\\
 -\dfrac{3\sin2\theta e^{i\phi}}{2\sqrt2} & 1-3\cos^2\theta & -\dfrac{3\sin2\theta e^{-i\phi}}{2\sqrt2} \\
 -\dfrac{3\sin^2\theta e^{i2\phi}}{2} & -\dfrac{3\sin2\theta e^{i\phi}}{2\sqrt2} & \dfrac{3\cos^2\theta-1}{2}
 \end{pmatrix},
\end{gathered}
\]
where $\theta,\phi$ are angles between $\bf R$ and quantization axis. The total two-body Hamiltonian is given by 
\begin{equation}
    \hat{H}_{\text{2b}} = \sum_{i=1,2}\big(H_0^{(i)}+H_{\text{shift}}^{(i)}\big)+\hat{V}_{\text{dd}}.
    \label{eq:full_H}
\end{equation}

We would like to briefly discuss the attractive and repulsive character of the dipole–dipole interaction potential. In the simple case where only a single excited molecular channel is relevant, while all other channels are energetically well separated, there's no eigenstate that is repulsive in all angles. For example, the potential corresponding to $|ge_0\rangle_\text{s}$ is repulsive for $\theta = \pi/2$ but attractive for $\theta = 0$, as shown by the orange curve in Fig.~\ref{fig:2}(b) in the main text. However, if there exists more than one near degenerate excitation channels, this will lead to global repulsive potentials regardless of CG coefficients, as in the green curve in Fig.~\ref{fig:2}(b) in the main text, $\Lambda$ systems or multiple degenerate states.

\subsubsection{Inelastic collision rate}
For a given molecular potential, the inelastic collision probability can be calculated analytically using the Landau-Zener method or numerically by solving Lindblad equations. We present both methods here.

\textit{The Landau-Zener method}. We first diagonalize $H_{\text{shift}} + V_{\text{dd}}$ inside $3\times3$ excited subspace and find the energy crossing point with ground state. Then we calculate the eigenvector and potential gradient at this point. The Landau-Zener tunneling propobility is given by
\begin{equation}
    P_{\text{tunnel}} = 1 - \exp\left(-\frac{2\pi\hbar\Omega^2}{v|s_+|}\right),
    \label{eq:tunnel_landau}
\end{equation}
where $v = \frac{\mathrm{d}R}{\mathrm{d}t}$, $s_+ = \frac{\mathrm{d}E(R)}{\mathrm{d}R}$, and $\Omega$ is the off-diagonal coupling strength. The atomic speed $v$ is estimated using the initial temperature of atoms in tweezers, which is set to $15~\mu$K based on experimental measurements. Note that this equation is only applicable for a single crossing point.

For the inelastic collision process, we assume head-on collisions and isotropic distribution of atoms, so the mean inelastic collision probability is 
\begin{equation}
    P_\text{ic} = \frac{1}{4\pi}\int 2P_{\text{tunnel}}(\theta,\phi)\big(1-P_{\text{tunnel}}(\theta, \phi)\big)\sin \theta \mathrm{d}\theta\mathrm{d}\phi.
    \label{eq:single_channel}
\end{equation}

This equation comes from one adiabatic and one non-adiabatic collisions and the factor of two comes from the two possibilities shown by the green and purple arrows in Fig.~\ref{fig:2}(a) in the main text. 

\textit{Lindblad equations}. In the presence of zero or more than one crossing points, Eq.~\ref{eq:tunnel_landau} is no longer accurate. For example, a noticeable loading enhancement [$P_\text{loading} = 66.4(7)\%$] is still observed in  Fig.~\ref{fig:7}(b) when there is no cross point (Fig.~\ref{fig:7}(d)). This behavior lies beyond the validity of the Landau–Zener tunneling framework. Instead, we determine the scattering probability by explicitly solving a Lindblad master equation.

The internal $4\times4$ Hamiltonian is given in Eq.~\ref{eq:full_H} and the motional degrees of freedom are treated classically. The spontaneous emission from the $^3P_1$ excited state to $^1S_0$ ground state is considered, while all other decoherence mechanisms are neglected. The system is initially prepared in the $\ket{g,g}$ state at a very large interatomic distance. The atoms then move toward each other with a relative velocity $v$, which is assumed to remain constant throughout the entire evolution.

To quantify the inelastic scattering associated with this process, we extract the minimum value of the $\ket{g,g}$ probability during the scattering process and identify it as the effective tunneling probability $P_{\mathrm{tunnel}}$. Then $P_\text{ic}$ can be calculated using Eq.~\ref{eq:single_channel}.

We note that the model is designed to work at $\hbar\Delta \in [U_0,2U_0]$ where loading enhancement is most significant. Outside this region, there are other effects that impair the enhanced loading, such as single-atom resonance heating ($\hbar\Delta \approx 0$) and the two-to-zero loss ($\hbar\Delta >2 U_0$).

\section{Enhanced loading with partially repulsive potentials}

\label{app:enhanced_loading}
The highest loading efficiency shown in Fig.~\ref{fig:1} in the main text is achieved with a $2.3$~G moderate magnetic field and $\pi$ polarization. 
Under this magnetic field, the $|^{3}\text{P}_1,m_J=-1\rangle$ state is slightly below the $|^{3}\text{P}_1,m_J=0\rangle$ cooling state, and therefore perturbs the red-detuned light-assisted collision process. 

\begin{figure}[h] 
    \centering
    \includegraphics[width=0.5\linewidth]
    {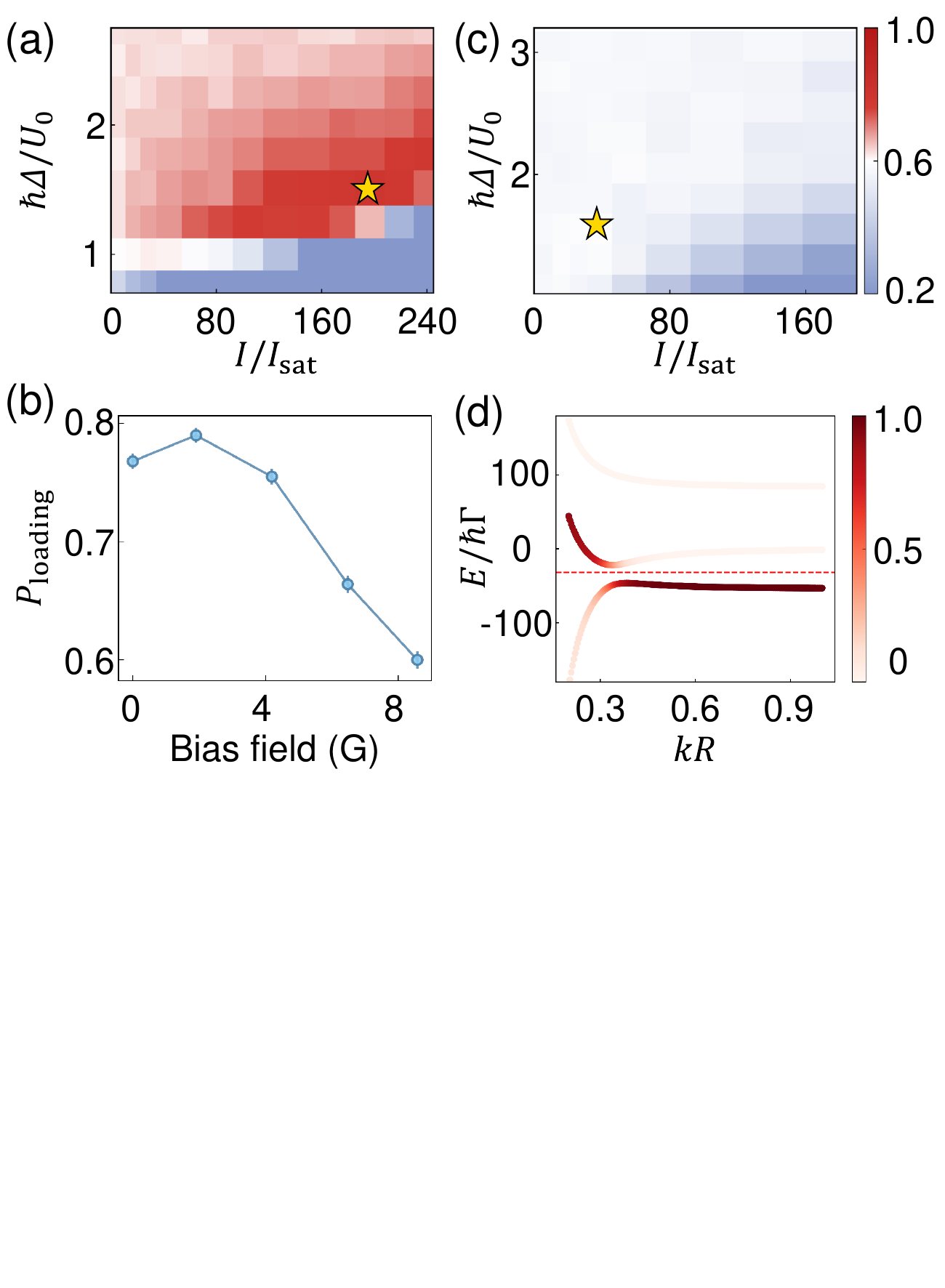}
    \caption{(a) The loading probability using $\pi$-polarized light under zero magnetic field. The detuning is referenced to the $\ket{e_0}$ state. (b) Loading probability as a function of the bias field for $\sigma^-$ light. (c) The loading probability using $\sigma^-$ light under a 8.6 G magnetic field. The detuning is referenced to the $\ket{e_{-1}}$ state. (d), Simulated potential energy curves at a collision angle of $\theta = 20^\circ$, with $B=6$~G. The color represents the relative coupling strength of $\sigma_-$ light to these states. Here $k = 2\pi/(556~\mathrm{nm})$. The dashed red line indicates the detuning of one trap depth to the $\ket{e_{-1}}$ state.} 
    \label{fig:7}
\end{figure}

To further investigate this effect, we provide more enhanced loading experiments with different potentials. We scan the detuning and intensity of $\pi$ polarized light under zero magnetic field (Fig.~\ref{fig:7}(a)). The highest loading efficiency reaches 79.9(6)\%. Under zero magnetic field, the tensor light shift for the $m_J = \pm1$ sublevels is smaller than the trap depth, while the collision light is detuned by more than one trap depth. As a result, the $\pi$-light partially couples to the globally repulse potentials, effectively increasing the loading efficiency.

We further investigate the enhancement effect with pure $\sigma^{-}$-light under magnetic field, which couples to a potential that is repulsive around $\theta=0$ and $180^\circ$ (Fig.~\ref{fig:7}(b)). At low field, the magnetic sublevels are not fully separated, yielding a modest increase in loading efficiency. However, when the magnetic field exceeds 1.9~G, the efficiency decreases with increasing field, eventually approaching the baseline of approximately 0.6 achieved with PA alone. To confirm the behavior of $\sigma^{-}$-light at high field, we perform a two-dimensional scan of the collision light parameters at 8.6~G (Fig.~\ref{fig:7}(c)) and no enhancement is observed. This is because the $\sigma^-$ light couples to the lowest single-atom energy level, and the dipole-dipole interaction opens a gap above that level, eliminating the repulsive part of the potential (Fig.~\ref{fig:7}(d)). 

\section{$^{174}$Yb gate scheme}
\label{qubit}
We propose encoding qubits in the ground $^1\text{S}_0$ and clock $^3\text{P}_0$ states of $^{174}$Yb atoms. Qubit storage requires a magic wavelength trap at 759 nm where available laser power is 30 W, smaller than 532 nm lasers. This is not a limit for the array size as qubit storage can be achieved with a much shallower trap as long as the atoms are sufficiently cold. In fact, the imaging and loading processes shown above are indeed the most power consuming part in the entire sequence~\cite{kaufman2022PRX-enhanceloading,Atomcomputing2025prx-171gsnqubit,jeff2023nature-erasure}. Atoms can be loaded and imaged in 532 nm tweezers and then transport to 759 nm tweezers for quantum computation.

Here we present a quantitative power-budget estimate for implementing a large-scale 759-nm magic tweezer array in the proposed clock-state architecture. As a benchmark, Ref.~\cite{kaufman2023PRX-omg} demonstrated clock-qubit operation in $^{171}\text{Yb}$ using 759-nm magic tweezers with a trap depth of $230~\text{kHz}$, corresponding to $11~\mu\text{K}$. Since the trapping properties of the $^1S_0$ and $^3P_0$ clock states are essentially isotope independent, this trap depth provides a useful reference for the proposed $^{174}\text{Yb}$ clock-state architecture. Using the trap-depth relation $U/h = \frac{2|\alpha|P}{\pi w^2},$
with the calculated 759 nm polarizability $\alpha = -8.4~\text{Hz}/(\text{W}/\text{cm}^2)$, the diffraction-limited waist $w = \lambda/(2 \text{NA}) = 759~\text{nm}/(2 \times 0.6) \approx 0.63~\mu\text{m}$, and an overall optical efficiency of $37\%$, dominated by the SLM diffraction efficiency reported in Ref.~\cite{jeff2024optica-localgate}, we estimate that a $230~\text{kHz}$ trap depth requires approximately $0.47~\text{mW}$ of laser output power per tweezer.

Therefore, a $10^4$-site 759 nm magic tweezer array would require approximately $4.7~\text{W}$ of laser output power. For the available power, commercial CW 759 nm laser systems currently provide up to $30~\text{W}$ output power, as available from Precilaser. Therefore, the available 759 nm laser power is not expected to be a limiting factor for clock-state storage at the demonstrated array scale.

For single-qubit gates, the fidelity of the optical qubit is 99.8\% which is lower than that of the hyperfine qubits~\cite{endres2025nature-6100atom} or the two qubit-gate fidelity estimated below. However, this is not a limitation for quantum computation as single-qubit gate errors is much more tolerable than two-qubit gate errors in quantum error correction~\cite{Preskill2002AIP-quantummemory,Andrew2021PRA-surfacecode}. 

For two-qubit gates, the main error sources that we considered are Rydberg state decay and unintended excitation to nearby overlapping Rydberg pair states. This assumption is justified by Ref.~\cite{jeff2025PRX-gatefidelity} where technical errors such as laser phase noise and Doppler shifts are smaller than the fundamental errors. We estimate $40~\mu$s lifetime of $|r\rangle$ using the measured value at $|\nu=54.28\rangle$ and lifetime $\propto \nu^3$~\cite{jeff2025PRX-gatefidelity}. 

\section{Distinction between alkali and alkaline-earth-like atom array platforms}

Here we present characteristics of alkali and alkaline-earth-like atoms for quantum computations. Alkali atoms, such as Rb and Cs, have a single valence electron, giving rise to ground-state hyperfine and Zeeman sublevels that are commonly used for qubit encoding. Single-qubit control is typically implemented with microwave or Raman fields, while entangling operations are mediated by two-photon excitation to Rydberg states~\cite{lukin2023nature-gatefidelity,saffman2025PRX-individualaddress}. In contrast, alkaline-earth-like atoms, such as Sr and Yb, have two valence electrons, giving rise to singlet and triplet electronic manifolds with narrow intercombination and clock transitions. This structure enables qubit encodings based on long-lived optical clock states~\cite{kaufman2023PRX-omg} or nuclear-spin states~\cite{kaufman2022PRX-enhanceloading,jeff2022PRX-UniversalGate,jeff2023nature-erasure}, as well as single-photon Rydberg excitation~\cite{endres2020np-AEAs,endres-nature2024benchmarking,jeff2023nature-erasure} and efficient quantum error correction using erasure conversion~\cite{jeff2022nc-erasure}. We list these distinctions in the following Table~\ref{tab:alkali_alkaline_earth}. 

\begin{table}[t]
    \centering
    \footnotesize
    \setlength{\tabcolsep}{2pt}
    \renewcommand{\arraystretch}{0.82}
    \caption{Distinction between alkali and alkaline-earth-like atom array platforms.}
    \label{tab:alkali_alkaline_earth}
    \begin{tabular*}{0.98\linewidth}{@{\extracolsep{\fill}}ccc}
        \hline\hline
        & \textbf{Alkali atom} & \textbf{Alkaline-earth-like atom} \\
        \hline
        \begin{tabular}{@{}c@{}}\textbf{Representative}\\[-3pt]\textbf{species}\end{tabular}
        & Rb, Cs
        & Yb, Sr \\
        \hline
        \textbf{Qubit encoding}
        & \begin{tabular}{@{}c@{}}
            Ground-state hyperfine qubit\\[-3pt]
            ~\cite{lukin2023nature-gatefidelity,saffman2025PRX-individualaddress}
          \end{tabular}
        & \begin{tabular}{@{}c@{}}
            Ground-state nuclear-spin qubit~\cite{kaufman2022PRX-enhanceloading,jeff2022PRX-UniversalGate}\\[-3pt]
            Metastable-state nuclear-spin qubit~\cite{jeff2023nature-erasure}\\[-3pt]
            Ground-metastable optical qubit~\cite{kaufman2023PRX-omg}
          \end{tabular} \\
        \hline
        \textbf{Coherence time}
        & \begin{tabular}{@{}c@{}}
            $T_2^* \sim 25\,\mathrm{ms}$, $T_2 \sim 12.6\,\mathrm{s}$~\cite{endres2025nature-6100atom}\\[-3pt]
            $T_2 \sim 20\,\mathrm{s}$~\cite{Zhang2024Optica-longcoherencetime}
          \end{tabular}
        & $T_2^* \sim 21\,\mathrm{s}$, $T_2 \sim 40\,\mathrm{s}$~\cite{Barnes2022nc-Yblongcoherencetime} \\
        \hline
        \begin{tabular}{@{}c@{}}\textbf{Rydberg}\\[-3pt]\textbf{excitation}\end{tabular}
        & Two-photon~\cite{lukin2023nature-gatefidelity,saffman2025PRX-individualaddress}
        & Single-photon~\cite{endres2020np-AEAs,endres-nature2024benchmarking,jeff2023nature-erasure} \\
        \hline
        \begin{tabular}{@{}c@{}}\textbf{Dominant 2Q gate}\\[-3pt]\textbf{error}\end{tabular}
        & Atom loss / Pauli error~\cite{lukin2023nature-gatefidelity,saffman2025PRX-individualaddress}
        & Erasure error~\cite{jeff2022nc-erasure} \\
        \hline\hline
    \end{tabular*}
\end{table}

\bibliography{main}